%% file: RJwrapper.tex
\documentclass[a4paper]{report}
\usepackage[utf8]{inputenc}
\usepackage[T1]{fontenc}
\usepackage{RJournal}
\usepackage{amsmath,amssymb,array}
\usepackage{booktabs}
\usepackage{graphicx}

\begin{document}

\sectionhead{Contributed research article}
\volume{XX}
\volnumber{YY}
\year{20ZZ}
\month{AAAA}

\begin{article}
  \input{hu-akande-wang}
\end{article}

\end{document}

%% file: hu-akande-wang.tex
\title{Multiple Imputation and Synthetic Data Generation with the R Package \pkg{NPBayesImputeCat}}
\author{by Jingchen Hu, Olanrewaju Akande and Quanli Wang}

\maketitle

\abstract{
	In many contexts, missing data and disclosure control are ubiquitous and challenging issues. In particular at statistical agencies, the respondent-level data they collect from surveys and censuses can suffer from high rates of missingness. Furthermore, agencies are obliged to protect respondents' privacy when publishing the collected data for public use. The \CRANpkg{NPBayesImputeCat} R package, introduced in this paper, provides routines to i) create multiple imputations for missing data, and ii) create synthetic data for statistical disclosure control, for multivariate categorical data, with or without structural zeros. We describe the Dirichlet process mixture of products of multinomial distributions model used in the package, and illustrate various uses of the package using data samples from the American Community Survey (ACS). We also compare results of the missing data imputation to the \CRANpkg{mice} R package and those of the synthetic data generation to the \CRANpkg{synthpop} R package.
}

\section{Introduction and background} \label{sec:intro}

\subsection{Multiple imputation for missing data}\label{sec:intro:MI}

Missing data problems arise in many statistical analyses. To impute missing values, multiple imputation, first proposed by \citet{Rubin1987}, has been widely adopted. This approach estimates predictive models based on the observed data, fills in missing values with draws from the predictive models, and produces multiple imputed and completed datasets. Data analysts then apply standard statistical analyses (e.g. a regression analysis) on each imputed dataset, and use appropriate combining rules to obtain valid point estimates and variance estimates \citep{Rubin1987}. 

As a brief review of the multiple imputation combining rules for missing data, let $q$ be the completed data estimator of some estimand of interest $Q$, and let $u$ be the estimator of the variance of $q$.  For $l=1, \dots, m$,  let $q^{(l)}$ and $u^{(l)}$ be the values of $q$ and $u$ in the $l$th completed dataset.  The multiple imputation estimate of $Q$ is equal to $\bar{q}_m = \sum_{l=1}^m q^{(l)}/m$, and the estimated variance  associated with  $\bar{q}_m$ is equal to $T_m = (1 + 1/m)b_m + \bar{u}_m$ , where $b_m = \sum_{l=1}^m (q^{(l)} - \bar{q}_m)^2/(m-1)$ and $\bar{u}_m = \sum_{l=1}^m u^{(l)}/m$.  Inferences for $Q$ are based on $(\bar{q}_m - Q) \sim t_{v}(0, T_m)$, where $t_{v}$ is a $t$-distribution with $v = (m-1)(1 + \bar{u}_m/ [(1+1/m) b_m])^2$ degrees of freedom.

Multiple imputation by chained equations (MICE, \citet{MICE}) remains the most popular method for generating multiple completed datasets after multiple imputation. Under MICE, one specifies univariate conditional models separately for each variable, usually using generalized linear models (GLMs) or classification and regression trees (CART \citet{breiman:1984, burgreit10}), and then iteratively samples plausible predicted values from the sequence of conditional models . For implementing MICE in R, most analysts use the \CRANpkg{mice} package. For an in-depth review of the MICE algorithm, see \citet{MICE}. For more details and reviews, see \citet{Rubin1996, HarelZhou2007, ReiterRaghu2007JASA}.

\subsection{Synthetic data for statistical disclosure control}\label{sec:intro:SD}

Statistical agencies regularly collect information from surveys and censuses, and make such information publicly available for various purposes, including research and policy making. In numerous countries around the world, statistical agencies are legally obliged to protect respondents' privacy when making these information available to the public. Statistical disclosure control (SDC) is the collection of techniques applied to confidential data before public release for privacy protection. Popular SDC techniques for tabular data including cell suppression and adding noise, and popular SDC techniques for respondent-level data (also known as microdata) include swapping, adding noise, and aggregation. \citet{SDC2012} provides a comprehensive review of SDC techniques and applications.

The multiple imputation methodology has been generalized to SDC. One approach to facilitating microdata release is to provide synthetic data. First proposed by \citet{Little1993} and \citet{Rubin1993}, the synthetic data approach estimates predictive models based on the original, confidential data, simulates synthetic values with draws from the predictive models, and produces multiple synthetic datasets. Data analysts then apply standard statistical analyses (e.g. a regression analysis) on each synthetic dataset, and use appropriate combining rules (different from those in multiple imputation) to obtain valid point estimates and variance estimates \citep{ReiterRaghu2007JASA, Drechsler2011book}. Moreover, synthetic data comes in two flavors: fully synthetic data \citep{Rubin1993} where every variable is deemed sensitive and therefore synthesized, and partially synthetic data \citep{Little1993} where only a subset of variables is deemed sensitive and synthesized, while the remaining variables are un-synthesized. Statistical agencies can choose between these two approaches depending on their protection goals, and subsequent analyses also differ. 

When dealing with fully synthetic data,  $\bar{q}_m$ estimates $Q$ as in the multiple imputation setting, but the estimated variance  associated with  $\bar{q}_m$ becomes $T_f = (1 + 1/m)b_m - \bar{u}_m$ , where $b_m$ and $\bar{u}_m$ are defined as in previous section on multiple imputation. Inferences for $Q$ are now based on $(\bar{q}_m - Q) \sim t_{v}(0, T_f)$, where the degrees of freedom is $v_f = (m-1)(1-m\bar{u}_m/((m+1)b_m))^2$.

For partially synthetic data,  $\bar{q}_m$ still estimates $Q$ but the estimated variance  associated with  $\bar{q}_m$ is $T_p = b_m/m + \bar{u}_m$ , where $b_m$ and $\bar{u}_m$ are defined as in the multiple imputation setting. Inferences for $Q$ are based on $(\bar{q}_m - Q) \sim t_{v}(0, T_p)$, where the degrees of freedom is $v_p = (m-1)(1 + \bar{u}_m/ [ b_m/m])^2$.

For synthetic data with R, \CRANpkg{synthpop} provides synthetic data generated by drawing from conditional distributions fitted to the confidential data. The conditional distributions are estimated by models chosen by the user, whose choices include parametric or CART models. For more details and reviews of synthetic data for statistical disclosure control, see \citet{Drechsler2011book}. 

\subsection{Structural zeros}\label{sec:intro:structuralzeros}

An important feature of survey data is the existence of structural zeros, which are combinations of variables with probability zero. For example, in the combinations of variables of vital signs, there should not exist a deceased patient with a pulse. For household survey, in the combinations of variables of relationship and age, there should not exist a household where a son is older than his biological father. As another example, if a dataset contains information of a record's age and educational attainment in the forms of categorical variables, there can be no record having the combination of being younger than 5 and having a doctorate degree.

In survey data with many variables, cross tabulations of variables could result in sparse tables, which contain non-structural zeros (combinations that are possible but happen not to exist in the particular dataset) and structural zeros (combinations that are simply impossible). To deal with structural zeros, many advanced statistical models are designed to assign zero probability for every impossible combination, which is a challenging task.

\subsection{What NPBayesImputeCat does}\label{sec:intro:pkg}

The \CRANpkg{NPBayesImputeCat} package specializes in estimating and performing multiple imputation and synthetic data generation for multivariate categorical data. Unlike \CRANpkg{mice} and \CRANpkg{synthpop}, both of which specifies conditional models, the \CRANpkg{NPBayesImputeCat} implements the Dirichlet process mixture of products of multinomial distributions (DPMPM), which specifies a joint latent class model on multivariate categorical variables. It uses Dirichlet process (DP) priors to allow effective clustering of the observations. Therefore, the \CRANpkg{NPBayesImputeCat} packages adds to the tools of imputation and synthesis, where a joint model might be more suitable than a series of conditional models for multivariate categorical data.

\CRANpkg{NPBayesImputeCat} also allows imputation with structural zeros. It therefore helps filling an important gap in missing data imputation techniques, as currently available R packages do not facilitate imputation with structural zeros, and users might have to do post-processing, such as rejection samplin,g to delete generated but impossible cases.

For multiple imputation, the \CRANpkg{NPBayesImputeCat} package allows data with and without structural zeros. For synthetic data, currently the package only allows data without structural zeros.

\subsection{The structure of this paper}\label{sec:intro:structure}

The rest of the paper is organized as follows. We first introduce the joint latent class models for multivariate categorical data that the \CRANpkg{NPBayesImputeCat} package applies, that is, the DPMPM model. 
In addition, we review applications of multiple imputation and synthetic data generation using the DPMPM in the literature. Next, we introduce the sample datasets from the American Community Survey (ACS) to be used in the demonstrations, and provide illustrations for both multiple imputation and synthetic data generation using the \CRANpkg{NPBayesImputeCat} package while comparing to other existing R packages. The paper concludes with summary and discussion.

\section{The DPMPM model} \label{sec:DPMPM}

Proposed by \citet{DunsonXing2009JASA}, the DPMPM is a Bayesian latent class model developed for multivariate categorical data. To allow for effective clustering of the observations based on all categorical variables, DP priors are specified for the mixture probabilities and multinomial probability vectors of the categorical data. The DPMPM has been shown to capture the complex dependencies in multivariate categorical data, while being computationally efficient. In addition, it empowers the data to select the number of latent classes to be used in the model estimation. The model has also been extended to account for structural zeros in categorical data \citep{ManriqueReiter2014JCGS}. 

The \CRANpkg{NPBayesImputeCat} package includes two versions of the DPMPM: i) DPMPM without structural zeros, and ii) DPMPM with structural zeros. In this section, we introduce the details of both versions and review previous work on using the DPMPM for multiple imputation and synthetic data.

\subsection{DPMPM without structural zeros} \label{sec:DPMPM:nozeros}

Our review of the DPMPM without structural zeros closely follows the review in \citet{HuHoshino2018PSD}. 
Consider a sample $\mathbf{X}$ consisting of $n$ records, where each $i$th record, with $i = 1, \ldots, n$, has $p$ unordered categorical variables. The basic assumption of the DPMPM is that every record $\mathbf{X}_i = (X_{i1}, \cdots, X_{ip})$ belongs to one of $K$ underlying unobserved/latent classes. Given the latent class assignment $z_i$ of record $i$, as in Equation (\ref{lca-z}), each variable $X_{ij}$ independently follows a multinomial distribution, as in Equation (\ref{lca-x}), where $d_j$ is the number of categories of variable $j$, and $j = 1, \ldots, p$. 
\begin{eqnarray}
	\label{lca-x} X_{ij} \mid z_i, \theta &\overset{ind}{\sim}& \textrm{Multinomial}(\theta_{z_i1}^{(j)}, \dots,
	\theta_{z_i d_j}^{(j)}; 1) \,\,\,\, \forall i, j\\
	\label{lca-z} z_i \mid \pi &\sim& \textrm{Multinomial}(\pi_1, \dots, \pi_K; 1) \,\,\,\, \forall i,
\end{eqnarray}

The marginal probability $Pr(X_{i1} = x_{i1}, \cdots, X_{ip} = x_{ip} \mid \pi, \theta)$ can be expressed as averaging over the latent classes:
\begin{equation}
	Pr(X_{i1} = x_{i1}, \cdots, X_{ip} = x_{ip} \mid \pi, \theta) = \sum_{k=1}^{K} \pi_k \prod_{j=1}^{p} \theta_{kx_{ij}}^{(j)}.
	\label{joint1}
\end{equation}
As pointed out in \citet{SiReiter2013JEBS, HuReiterWang2014PSD, AkandeLiReiter2017TAS}, such averaging over latent classes results in dependence among the variables. Equation (\ref{joint1}) will also help illustrate the DPMPM with structural zeros in the next section.

The DPMPM clusters records with similar characteristics based on all $p$ variables. Relationships among all the variables are induced by integrating out the latent class assignment $z_i$. To empower the DPMPM to pick the effective number of occupied latent classes, the truncated stick-breaking representation \citep{Sethuraman1994SS} of the DP prior is used as in Equation (\ref{prior-pi}) through Equation (\ref{prior-theta}),
\begin{eqnarray}
	\label{modelprior}
	\label{prior-pi} \pi_k &=& V_{k}\prod_{l<k}(1-V_{l}) \,\,\,\, \textrm{for } k = 1, \dots, K\\
	\label{prior-V} V_{k}&\overset{iid}{\sim}& \textrm{Beta} (1,\alpha) \,\,\,\, \textrm{for } k = 1, \dots, K-1, \,\,\,\, V_K=1\\
	\label{prior-alpha} \alpha &\sim& \textrm{Gamma}(a_{\alpha}, b_{\alpha})\\
	\label{prior-theta} \boldsymbol{\theta}_{k}^{(j)}=(\theta_{k1}^{(j)},\dots,\theta_{kd_j}^{(j)})&\sim& \textrm{Dirichlet} (a_{1}^{(j)}, \dots,a_{d_{j}}^{(j)})  \,\,\,\, \textrm{for } j = 1, \dots, p, \ \ \ k = 1, \dots, K.
\end{eqnarray}
and a blocked Gibbs sampler is implemented for the Markov chain Monte Carlo (MCMC) sampling procedure \citep{IshwaranJames2001JASA, SiReiter2013JEBS, HuReiterWang2014PSD, AkandeLiReiter2017TAS, ManriqueHu2018JRSSA, DrechslerHu2018, HuSavitsky2018}. 

When used as an imputation engine, missing values are handled within the Gibbs sampler. As described in \citet{AkandeLiReiter2017TAS}, at one MCMC iteration $l$, one samples a value of the latent class indicator $z_i$ using Equation (\ref{lca-z}), given a draw of the parameters and observed data. In this iteration $l$, given the sampled $z_i$, one samples missing values using independent draws from Equation (\ref{lca-x}). This process is repeated for every missing value in the dataset in iteration $l$, obtaining one imputed dataset.

When used as a data synthesizer, the fully observed confidential dataset is used for model estimation through MCMC, and sensitive variable values are synthesized as an extra step at chosen MCMC iteration. For example, at MCMC iteration $l$, one samples a value of the latent class indicator $z_i$ using Equation (\ref{lca-z}). Given the sampled $z_i$, one samples synthetic values of sensitive variables using independent draws from Equation (\ref{lca-x}). This process is repeated for every record who has sensitive values to be synthesized, obtaining one synthetic dataset.

\subsection{DPMPM with structural zeros} \label{sec:DPMPM:zeros}

When structural zeros are present, we need to modify the likelihood to enforce zero probability for impossible combinations. That is, we need to truncate the support of the DPMPM. Following the general description in \citet{ManriqueReiter2014JCGS, ManriqueHu2018JRSSA}, let $\mathcal{C}$ represent all combinations of individuals, including impossible combinations; let $\mathcal{MCZ} \not\subseteq \mathcal{C}$ be the set of impossible combinations to be excluded. We restrict $\mathbf{X}$ to the set $\mathcal{C} \setminus \mathcal{MCZ}$, with $Pr(\mathbf{X} \in \mathcal{MCZ}) = 0$. The marginal probability in the DPMPM without structural zeros in Equation (\ref{joint1}) then becomes
\begin{equation}
	Pr(\mathbf{X}_i = \mathbf{x}_i \mid \pi, \theta, \mathcal{MCZ})  \propto I(\mathbf{X}_i \notin \mathcal{MCZ}) \sum_{k=1}^{K} \pi_k \prod_{j=1}^{p} \theta_{kx_{ij}}^{(j)}.
	\label{joint2}
\end{equation}
Let $\mathcal{X}^*$ be the sample that only contains possible combinations, we have the joint likelihood as
\begin{equation}
	p(\mathcal{X}^* \mid \pi, \theta, \mathcal{MCZ}) \propto \prod_{i=1}^{n} I(\mathbf{X}_i \notin \mathcal{MCZ}) \sum_{k=1}^{K} \pi_k \prod_{j=1}^{p} \theta_{kx_{ij}}^{(j)}.
	\label{DPMPMzeros_lik}
\end{equation}

To get the Gibbs sampler to work, we follow the general data augmentation technique proposed by \citet{ManriqueReiter2014JCGS}, and assume the existence of an observed sample $\mathcal{X}^0$ of unknown size $Nmis$, generated from the DPMPM without structural zeros (i.e. the unrestricted DPMPM). $\mathcal{X}^0$ only contains records that fall into $\mathcal{MCZ}$. 

The same set of DP priors in Equation (\ref{prior-pi}) through Equation (\ref{prior-theta}) are used in the DPMPM with structural zeros. In the Gibbs sampler, we keep the generated $\mathcal{X}^0$ and combine it with $\mathcal{X}^*$ when estimating the model parameters. For computational expedience, we set the upper bound of the number of observations, $Nmis$, that can be generated in $\mathcal{X}^0$, to be fixed at a large $Nmax$ at every iteration.  When used as either an imputation engine or a data synthesizer, missing values or synthetic data are generated from the truncated likelihood Equation (\ref{DPMPMzeros_lik}).

\subsection{Applications of DPMPM for multiple imputation}\label{sec:DPMPM:MI}
The DPMPM has been adapted as a multiple imputation engine to deal with missing values in categorical data. Some imputation applications have focused on the DPMPM without structural zeros, while others have dealt with the DPMPM with structural zeros.

Among the work on multiple imputation using the DPMPM without structural zeros, \citet{SiReiter2013JEBS} applied the DPMPM imputation model to impute missing background data (categorical) in the 2007 Trends in International Mathematics and Science Study (TIMSS). The 2007 TIMSS data contains 80 background variables on 90,505 students. Among the 80 categorical background variables, 68 have less than 10\% missing values, 6 variables have between 10\% and 30\% missing values, and 1 variable has more than 75\% missing values. 

\citet{AkandeLiReiter2017TAS} designed simulation studies using data from the American Community Survey (ACS), and compared the DPMPM imputation engine to two other widely used multiple imputation engines: i) chained equations using generalized linear models, and ii) chained equations using classification and regression trees (CART). From a population of 671,153 housing units and 35 categorical variables collected and cleaned from the 2012 ACS data, \citet{AkandeLiReiter2017TAS} performed repeated sampling and empirically compared the three multiple imputation models. 

Among the work on multiple imputation using the DPMPM with structural zeros, \citet{ManriqueReiter2014SM} followed the data augmentation approach \citet{ManriqueReiter2014JCGS}, and imputed missing data of repeated samples from the 5\% public use microdata sample from the 2000 United States Census for the state of New York, a population of 953,076 individuals and 10 categorical variables, with the number of levels ranging from 2 to 11.

Finally, \citet{Murray2018SS} provide an excellent review of practical and theoretical findings of multiple imputation research, and highlights the DPMPM imputation engine as a recent development.

\subsection{Applications of DPMPM for synthetic data}\label{sec:DPMPM:synthesis}

The DPMPM has also been used as a synthetic data generator to public release of useful and private microlevel categorical data. Some work focused on the DPMPM without structural zeros, while others dealt with synthetic data problems using the DPMPM with structural zeros.

Among the work on synthetic data generation using the DPMPM without structural zeros, \citet{HuReiterWang2014PSD} used the DPMPM to generate fully synthetic data for a subset of 10,000 individuals and 14 categorical variables from the 2012 ACS public use microdata sample for the state of North Carolina. \citet{DrechslerHu2018} generated partially synthetic data for a large-scale administrative data containing detailed geographic information in Germany.  \citet{HuSavitsky2018} also used the DPMPM to generate partially synthetic data for the Consumer Expenditure Surveys (CE) at the U.S. Bureau of Labor Statistics (BLS), disseminating detailed county-level geographic information.

Among the work on synthetic data generation using the DPMPM with structural zeros, \citet{ManriqueHu2018JRSSA} proposed a data augmentation approach, and generated fully synthetic data of repeated samples from the 5\% public use microdata from the 2000 United States Census for the state of California, a population of 1,690,642 records measured in 17 categorical variables, with the number of levels ranging from 2 to 11.

\section{Two ACS samples for illustrations} 
\label{sec:ACSsamples}

Before presenting detailed step-by-step illustrations to use the \CRANpkg{NPBayesImputeCat} package for multiple imputation and data synthesis applications, we introduce two samples from the 2016 1-year American Community Surveys (ACS), which will both be used for our illustrations. 

ACS sample 1, \file{ss16pusa\_sample\_zeros}, contains structural zeros. It will be used to illustrate how to perform multiple imputation and data synthesis tasks when structural zeros are present. ACS sample 2, \file{ss16pusa\_sample\_nozeros}, is a subset of ACS sample 1 and contains no structural zeros. It will be used to illustrate how to perform multiple imputation and data synthesis tasks when structural zeros are not present.

\subsection{ACS sample 1, with structural zeros}
\label{sec:ACSsamples:1}
\begin{table}[bth]
	\small
	\centering
	\caption{Variables used in ACS sample 1. The four table columns provide information on: variable name, simple description of the variable, the number of categories, and the details of the categories.}
	\label{tab:ACSvars1}
	\begin{tabular}{p{0.4in} p{1.2in} p{0.1in} p{3.3in}}
		\hline
		Variable &  Description & $\#$  & Category details \\ \hline
		AGEP &  Age &7 &16; 17; [18, 24]; [25, 35]; [36, 50]; [51, 70]; (70, )\\
		MAR & Marital status& 5& Married; Widowed; Divorced; Separated; Never married. \\
		SCHL & Education attainment&9& Up to K0; Some K12, no diploma; High school diploma or GED; Some college, no degree; Associate's degree; Bachelor's degree; Master's degree; Professional degree; Doctorate degree. \\
		SEX & Sex&2& Male; Female \\
		WKL & When last worked&3& Within the last 12 months; 1-5 years ago; Over 5 years ago or never worked. \\ \hline
	\end{tabular}
\end{table}

ACS sample 1 is a random sample of $n = 1,000$ observations on $p = 5$ variables. See Table \ref{tab:ACSvars1} for the data dictionary. The sample is saved as \file{ss16pusa\_sample\_zeros}, and it contains structural zeros: 8 combinations, all related to AGEP and SCHL variables, listed in Table \ref{tab:ACSzeros}. These 8 cases are derived from the original 2016 1-year ACS data (as the population).
\begin{table}[bth]
	\small
	\centering
	\caption{8 cases of structural zeros in the ACS sample. The two table columns include the index of each structural zeros case and simple description of the case itself.}
	\label{tab:ACSzeros}
	\begin{tabular}{c l | c l}
		\hline
		 $\#$ &  Description & $\#$   & Description \\ \hline
		1 &  AGEP = 16 \& SCHL = Bachelor's degree. & 5 &  AGEP = 17 \& SCHL = Bachelor's degree.\\
		2 &  AGEP = 16 \& SCHL = Doctorate degree. & 6 &  AGEP = 17 \& SCHL = Doctorate degree.\\
		3 &  AGEP = 16 \& SCHL = Master's degree. & 7 &  AGEP = 17 \& SCHL = Master's degree.  \\
		4 & AGEP = 16 \& SCHL = Professional degree. & 8 & AGEP = 17 \& SCHL = Professional degree. \\ \hline
	\end{tabular}
\end{table}

\subsection{ACS sample 2, without structural zeros}
\label{sec:ACSsamples:2}

To obtain a sample without structural zeros, we take a subset of ACS sample 1, where $n = 1,000$ and $p = 3$, dropping variables AGEP and SCHL to eliminate any structural zeros. This ACS sample 2 is saved as \file{ss16pusa\_sample\_nozeros}. See Table \ref{tab:ACSvars2} for the data dictionary.

\begin{table}[bth]
	\small
	\centering
	\caption{Variables used in ACS sample 2. The four table columns provide information on: variable name, simple description of the variable, the number of categories, and the details of the categories.}
	\label{tab:ACSvars2}
	\begin{tabular}{p{0.4in} p{1.2in} p{0.1in} p{3.3in}}
		\hline
		Variable &  Description & $\#$ & Category details \\ \hline
		MAR & Marital status& 5& Married; Widowed; Divorced; Separated; Never married. \\
		SEX & Sex&2& Male; Female \\
		WKL & When last worked&3& Within the last 12 months; 1-5 years ago; Over 5 years ago or never worked. \\ \hline
	\end{tabular}
\end{table}

\section{Missing data applications}  \label{sec:illustrationsMI}

To illustrate the applications of the \CRANpkg{NPBayesImputeCat} package to missing data, we introduce 30\% missingness for each variable in the ACS sample 1 and ACS sample 2 datasets, under the missing completely at random (MCAR) mechanism. 
The corresponding datasets (containing missing values) to ACS sample 1 and ACS sample 2 are saved as \file{ss16pusa\_sample\_zeros\_miss} and  \file{ss16pusa\_sample\_nozeros\_miss} respectively. The DPMPM imputation engine is designed to perform multiple imputations of categorical data that are missing at random (MAR)--and thus also data that are missing completely at random (MCAR).

\subsection{Multiple imputation for data without structural zeros}
\label{sec:illustrationsMI:nozeros}

We begin with  the imputation of the missing values in the ACS sample 2 with 30\% missingness, \file{ss16pusa\_sample\_nozeros\_miss}, where no structural zeros are present. In the next section, we demonstrate how to impute missing values for ACS sample 1 with 30\% missingness, \file{ss16pusa\_sample\_zeros\_miss}, where structural zeros are present.

For each sample, we also compare the performance of the DPMPM engine to the most popular multiple imputation method, MICE. We implement the latter using the \CRANpkg{mice} package in R. A brief review of \CRANpkg{mice} is included at the beginning of the paper. 

\subsubsection*{{\underline{Load the sample data}}}
First, we load the sample data, the ACS sample 2 with 30\% missingness, and make sure that all variables are unordered factors.
\begin{verbatim}
data("ss16pusa_sample_nozeros_miss")
X <- ss16pusa_sample_nozeros_miss
p <- ncol(X)
for (j in 1:p){
  X[,j] <- as.factor(X[,j])
}
\end{verbatim}

\subsection*{{\underline{Initialize the DPMPM imputation engine}}}
We use the \code{DPMPM\_nozeros\_imp} function to implement the DPMPM imputation engine without structural zeros. We first review the process for creating and initializing the DPMPM model using the \code{CreateModel} function, to enable analysts tune the number of mixture components through initial runs, before generating imputations using \code{DPMPM\_nozeros\_imp}. \code{CreateModel} is a wrapper function for creating an object of type “Lcm”. Lcm was implemented as an Rcpp module to expose the C++ implementation for our algorithm. Users can learn more about the \code{Lcm} class by typing \code{?`Lcm'}, which will bring up the R documentation for this class, including all methods and properties. 

\code{CreateModel} takes 7 arguments as input: 
\begin{enumerate}
	\item \code{X}, the original data with missing values
	\item \code{MCZ}, the data frame containing the structural zeros definitions - use \code{NULL} when structural zeros are not present
	\item \code{K}, the maximum number of mixture components (i.e. the maximum number of latent classes in the DPMPM imputation engine)
	\item \code{Nmax}, an upper truncation limit for the augmented sample size, that is, the maximum number of observations allowable in the augmented $\mathcal{X}^0$ - use \code{0} when structural zeros are not present
	\item \code{aalpha}, the hyper parameter $a_{\alpha}$ in stick-breaking prior distribution in Equation (\ref{prior-alpha})
	\item \code{balpha}, the hyper parameter $b_{\alpha}$ in stick-breaking prior distribution in Equation (\ref{prior-alpha})
	\item \code{seed}, the seed value.
\end{enumerate}

As a quick demonstration, we let \code{K} $= 30$, \code{aalpha} $=$ \code{balpha} $=$ 0.25, and \code{seed} $= 456$. The code below creates and initializes the DPMPM imputation engine without structural zeros for the data stored in \code{X}.
\begin{verbatim}
	model <- CreateModel(X = X,
                     MCZ = NULL,
                     K = 30,
                     Nmax = 0,
                     aalpha = 0.25,
                     balpha = 0.25,
                     seed = 456)
\end{verbatim}
Next, we run the \code{model} object for a set of user-specified numbers of burn-ins, MCMC iterations, and thinning. For example, to run the MCMC sampler for 5 iterations post 2 burn-ins, thin every 1 iteration, and print the output at each iteration, run the following code. 

\begin{verbatim}
	> model$Run(burnin = 2, 
          		 iter = 5, 
          		 thinning = 1, 
         		  silent = FALSE)
Initializing...
Run model without structural zeros.
iter = 0  kstar = 30 alpha = 1 Nmis = 0
iter = 0  kstar = 30 alpha = 7.81552 Nmis = 0
iter = 1  kstar = 30 alpha = 6.6941 Nmis = 0
iter = 2  kstar = 30 alpha = 4.60622 Nmis = 0
iter = 3  kstar = 30 alpha = 5.67409 Nmis = 0
\end{verbatim}
Here, we show the first few lines of the output. The output prints out the iteration index as \code{iter}, the value of occupied mixture components or latent classes as \code{kstar}, posterior estimates of $\alpha$ (the concentration parameter in stick-breaking prior distribution in Equation (\ref{prior-alpha})) as \code{alpha}, and the size of the augmented sample as \code{Nmis}. In our demonstration, \code{Nmis} is always 0, as the size of the augmented sample is 0 when there are no structural zeros. 

It is important to keep track of the value of \code{kstar}, as the \CRANpkg{NPBayesImputeCat} package uses the truncated stick-breaking representation of the DP prior \citep{Sethuraman1994SS}. If the value of \code{kstar} is always \code{K}, the maximum number of mixture components, we should re-run the DPMPM model by specifying a larger value of \code{K}, to allow large enough number of mixture components to cluster the observations. For additional details on setting \code{K}, see \citet{HuReiterWang2014PSD, AkandeLiReiter2017TAS}.

The above initial run seems to suggest that the estimation uses almost all latent classes (\code{kstar} is close or is 30, which is what the maximum number of latent classes \code{K} set to). It is therefore prudent to increase the value of \code{K} when executing the \code{CreateModel} command, for example:
\begin{verbatim}
	> model <- CreateModel(X = X,
                     		 MCZ = NULL,
                     		 K = 80,
                     		 Nmax = 0,
                     		 aalpha = 0.25,
                     		 balpha = 0.25,
                     		 seed = 456)
	> model$Run(burnin = 2, 
            iter = 5, 
            thinning = 1, 
            silent = FALSE)
Initializing...
Run model without structural zeros.
iter = 0  kstar = 80 alpha = 1 Nmis = 0
iter = 0  kstar = 78 alpha = 16.4979 Nmis = 0
iter = 1  kstar = 77 alpha = 17.281 Nmis = 0
iter = 2  kstar = 75 alpha = 24.4488 Nmis = 0
iter = 3  kstar = 79 alpha = 26.1196 Nmis = 0
\end{verbatim}  
Again, we only show the first few lines of the output. This time, setting \code{K} equal to 80 seems sufficiently large. Users should keep track of the value of \code{kstar} for the entire run, and adjust \code{K} accordingly.

To diagnose convergence of parameters in the Gibbs sampler, one can use the \code{EnableTracer} option before running the sampler, to track certain parameters. Examples of keeping posterior samples of \code{alpha} and \code{kstar} to access convergence are included in the accompanying R file.
%
%
%
%
%

\subsection*{{\underline{Generate the imputed datasets}}}
After setting \code{K} based on the initial runs, we now run the DPMPM imputation engine without structural zeros to create $m$ imputed datasets. The function \code{DPMPM\_nozeros\_imp} takes 10 arguments as input:
\begin{enumerate}
	\item \code{X}, the original data with missing values
	\item \code{nrun}, the number of MCMC iterations
	\item \code{burn}, the number of burn-in
	\item \code{thin}, the number of thinning
	\item \code{K}, the maximum number of mixture components (i.e. the maximum number of latent classes in the DPMPM imputation engine) \item \code{aalpha}, the hyper parameter $a_{\alpha}$ in stick-breaking prior distribution in Equation (\ref{prior-alpha})
	\item \code{balpha}, the hyper parameter $b_{\alpha}$ in stick-breaking prior distribution in Equation (\ref{prior-alpha})
	\item \code{m}, the number of imputations
	\item \code{seed}, the seed value
	\item \code{silent}, default to TRUE. Set this parameter to FALSE if more iteration info are to be printed.
\end{enumerate}
The output of \code{DPMPM\_nozeros\_imp} is a list containing: 

\begin{enumerate}
	\item \code{impdata}, the list of $m$ imputed datasets
	\item \code{origdata}, the original data \code{X}
	\item \code{alpha}, the saved posterior draws of $\alpha$, which can be used to check MCMC convergence
	\item \code{kstar}, the saved numbers of occupied mixture components, which can be used to check MCMC convergence and track whether the upper bound \code{K} is set large enough. 
\end{enumerate}
To run the \code{DPMPM\_nozeros\_imp} function to impute missing data for ACS sample 2 with 30\% missingness, we run the code below. For this demonstration, we set \code{nrun} to 10000, \code{burn} to 5000, \code{thin} to 50, \code{K} to 80,  both \code{aalpha} and \code{balpha} 0.25, and \code{m} to 10. Finally, we set the \code{seed} to 211.
\begin{verbatim}
	m <- 10
	Imp_DPMPM <- DPMPM_nozeros_imp(X = X,
                               nrun = 10000,
                               burn = 5000,
                               thin = 50,
                               K = 80,
                               aalpha = 0.25,
                               balpha = 0.25,
                               m = m,
                               seed = 211,
                               silent = TRUE)
\end{verbatim}
The printed output from each iteration are omitted here. For a quick diagnostic check on whether the upper bound \code{K} is set large enough, we can use the \code{kstar\_MCMCdiag} function which takes the following input arguments:
\begin{enumerate}
	\item \code{kstar}, the vector output of kstar from running the DPMPM model
	\item \code{nrun}, the number of MCMC iterations used in running the DPMPM model
	\item \code{burn}, the number of burn-in iterations used in running the DPMPM model
	\item \code{thin}, the number of thinning used in running the DPMPM model.
\end{enumerate}

Its output a list of two MCMC diagnostics figures:

\begin{enumerate}
	\item \code{Traceplot}, the traceplot of kstar post burn-in and thinning
	\item \code{Autocorrplot}, the autocorrelation plot of kstar post burn-in and thinning
\end{enumerate}

\begin{figure}[t!]
	\centering
	\includegraphics[width=0.47\linewidth]{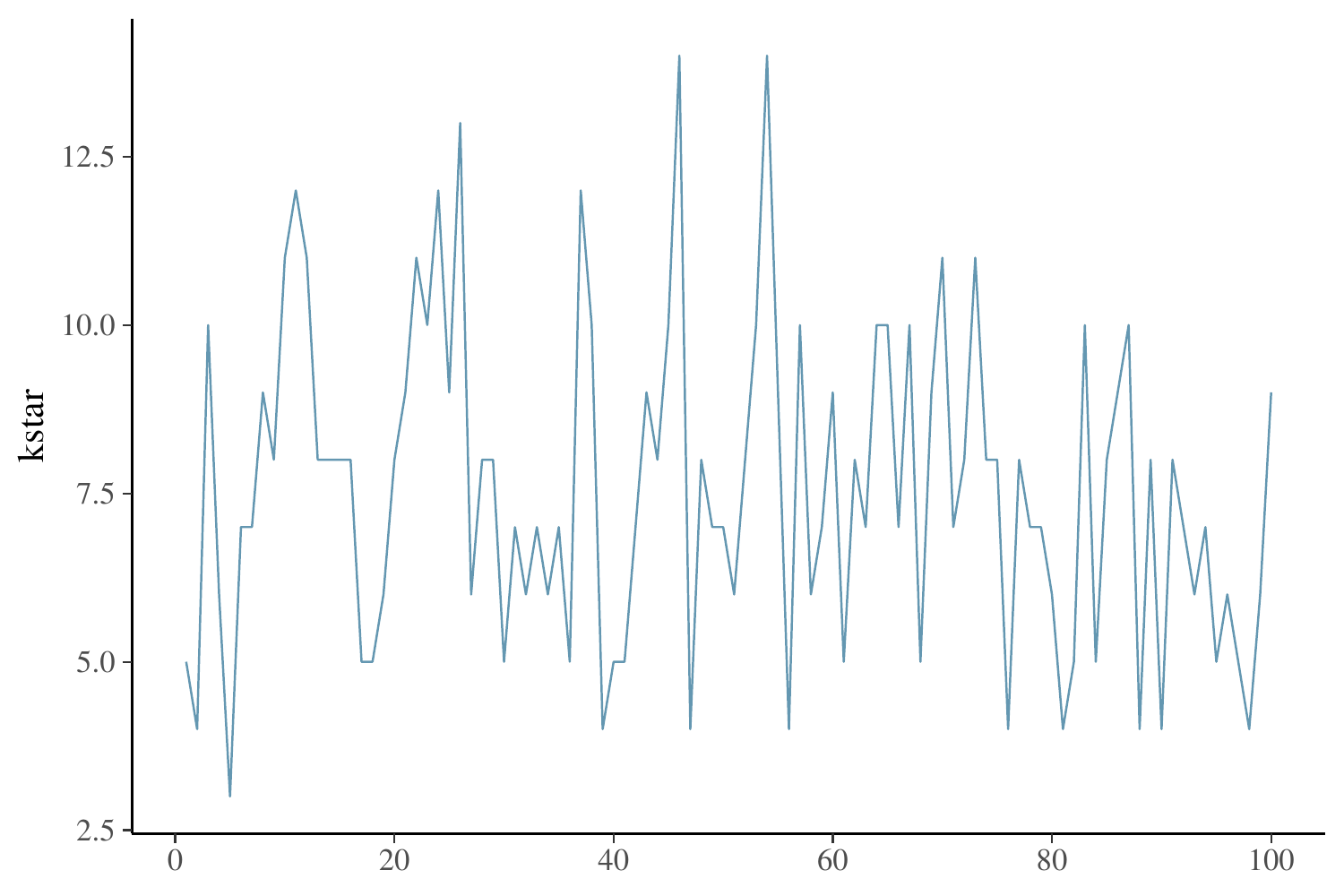}
	\caption{\label{kstar_traceplot}Traceplot of the thinned kstar values after burn-in. It shows little stickiness (only a 100 samples). The mean  of kstar is 7.5, with range from 2.5 to 15. This suggests that setting $K = 80$ should be sufficient and we could consider setting a smaller $K$ for an even faster computation time.}
\end{figure}

\begin{figure}[t!]
	\centering
	\includegraphics[width=0.47\linewidth]{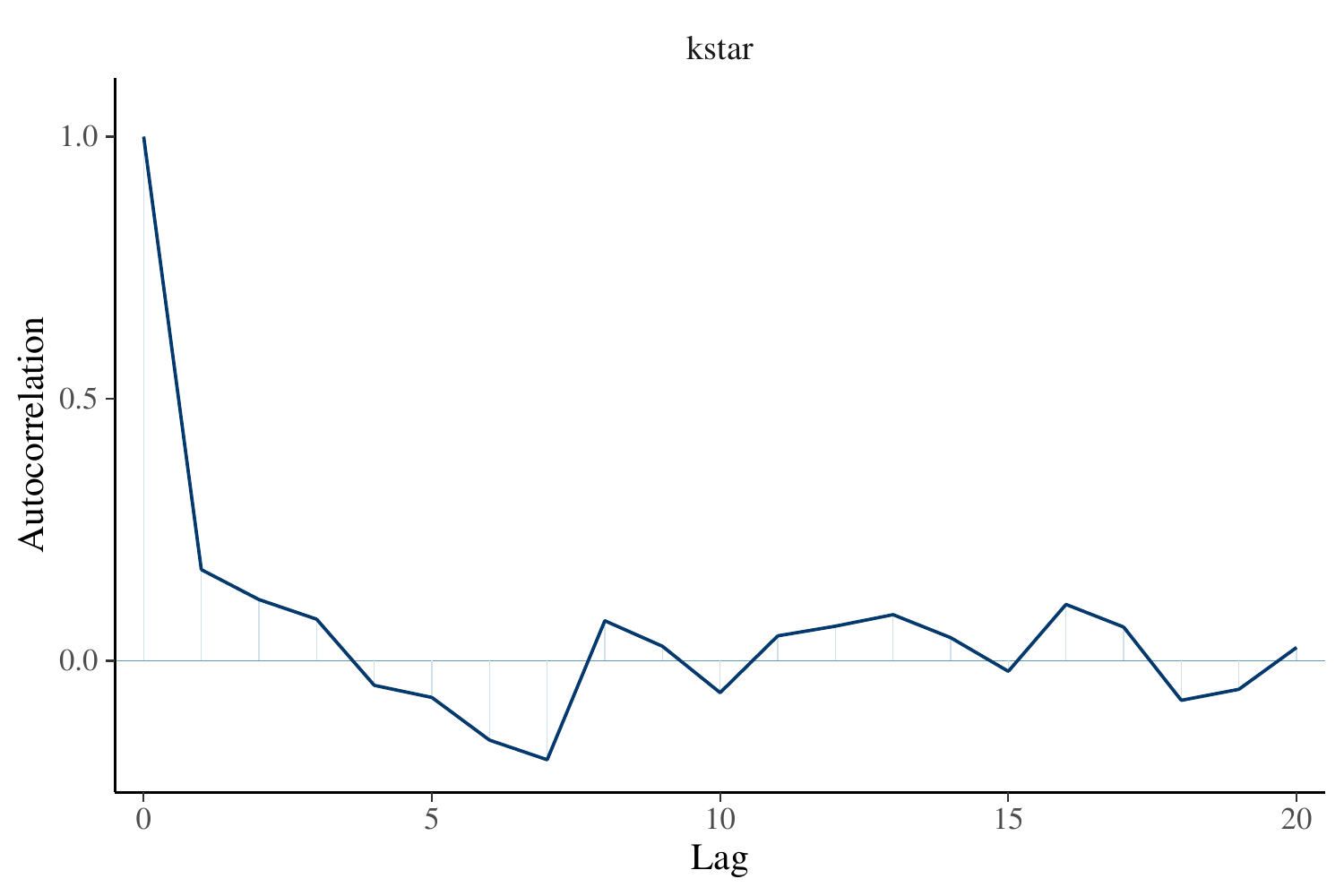}
	\caption{\label{kstar_ACF}Autocorrelation plot of the thinned kstar values after burn-in. It shows a sharp drop in autocorrelation after lag 1, indicating no convergence issues after performing such MCMC diagnostics. }
\end{figure}

We first load the \CRANpkg{bayesplot} package before using the \code{kstar\_MCMCdiag} function.
\begin{verbatim}
library(bayesplot)
kstar_MCMCdiag(kstar = Imp_DPMPM$kstar,
                nrun = 10000,
                burn = 5000,
                thin = 50)
\end{verbatim}
Figure \ref{kstar_traceplot} shows the traceplot of \code{kstar} value after burn-in and thinning. It indicates no convergence issues of the MCMC chain. Moreover, it suggests choosing a smaller \code{K} value if we want to achieve an even faster computation time. Figure \ref{kstar_ACF} presents its autocorrelation function plot which also indicates no convergence issues.


To access the imputed datasets one at a time, we do the following.
\begin{verbatim}
	impdata1 <- Imp_DPMPM$impdata[[1]] #for the first imputed dataset
\end{verbatim}
Analysts then can compute sample estimates for estimands of interest in each imputed dataset, and combine them using the combining rules.

Before demonstrating how to do so, we first use the \CRANpkg{mice} package to also generate imputations for the same dataset. We do so to facilitate comparisons between results based on the DPMPM model and MICE.  The following code runs the MICE algorithm on the same data using the default options in MICE for all the arguments, except \code{m}, which is set to 10 to be consistent with the implementations of the DPMPM engine. The code also reshapes the output of the MICE algorithm so that we are able to use some of the utility functions in \CRANpkg{NPBayesImputeCat} on the imputed datasets. For more details on the implementations of \CRANpkg{mice}, see \citet{MICE}.
\begin{verbatim}
	library(mice)
m <- 10
Imp_MICE <- mice(data = X, 
                 m = m,
                 defaultMethod = c("norm", "logreg", "polyreg", "polr"),
                 print = F, 
                 seed = 342)
                 
#Reshape the list of imputed datasets
Imp_MICE_reshape <- NULL
Imp_MICE_reshape$impdata <- lapply(1:m,function(x) x <- X)
col_names <- names(Imp_MICE$imp)
for (l in 1:m){
  for(j in col_names){
    na_index_j <- which(is.na((Imp_MICE_reshape$impdata[[l]])[,j])==TRUE)
    Imp_MICE_reshape$impdata[[l]][na_index_j,j] <- Imp_MICE$imp[[j]][[l]]
  }
}
\end{verbatim}
With the \code{Imp\_DPMPM}, \code{Imp\_MICE}, \code{Imp\_MICE\_reshape} objects, we now demonstrate how to assess the quality of the imputations for the two methods, and also use the combining rules for valid inferences from multiple imputed datasets.

\subsection*{{\underline{Assess quality of the imputations}}}
\begin{figure}[t!]
	\centering
	\includegraphics[width=0.7\linewidth]{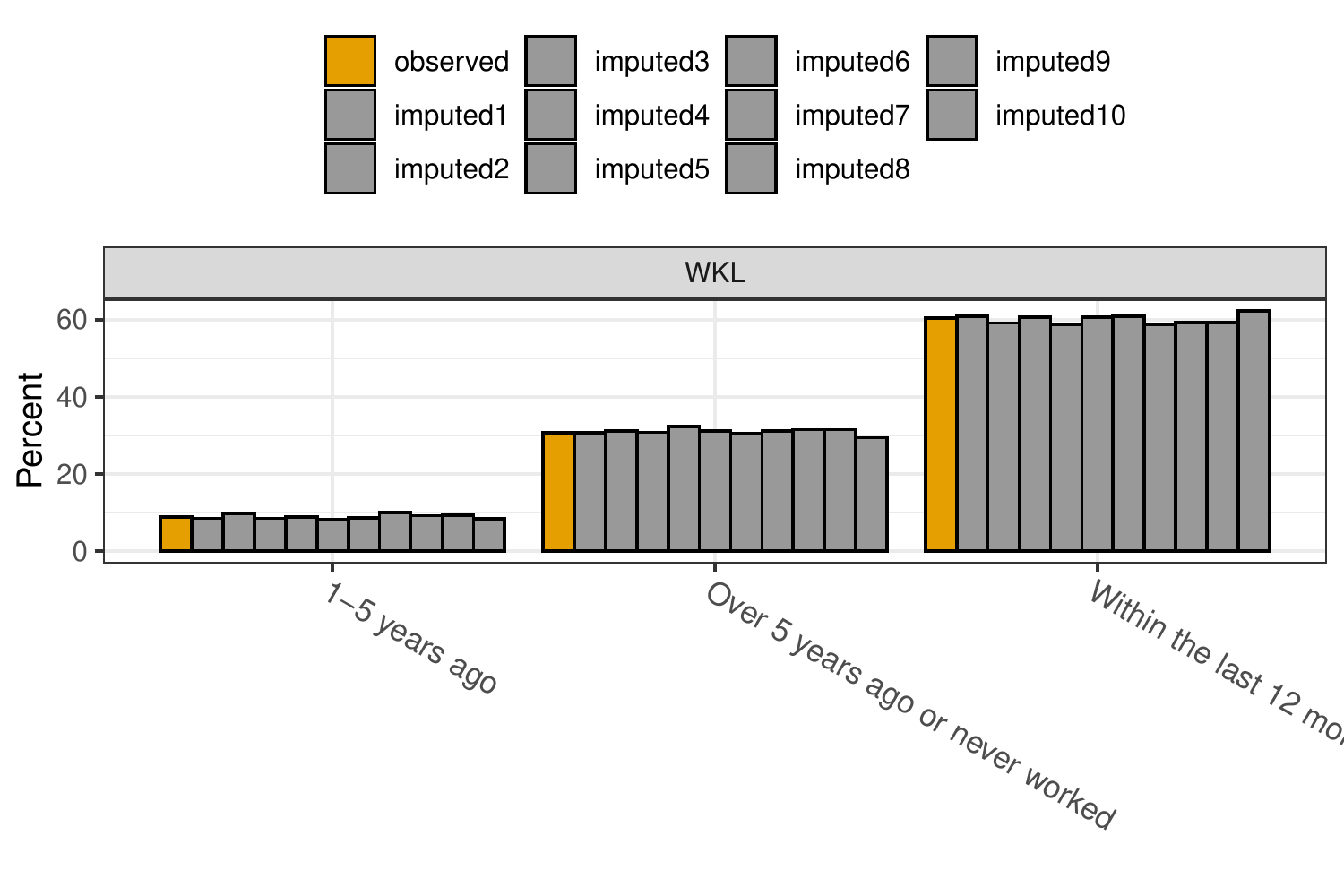}
	\caption{\label{DPMPMresults1}Marginal distribution of WKL from the observed data and each imputed dataset, using DPMPM. Barplots of the observed (yellow) and the 10 imputed (grey) are shown for the three levels of WKL. There is some variability across the imputed datasets. Overall they resemble the observed well.}
\end{figure}

\begin{figure}[t!]
	\centering
	\includegraphics[width=0.7\linewidth]{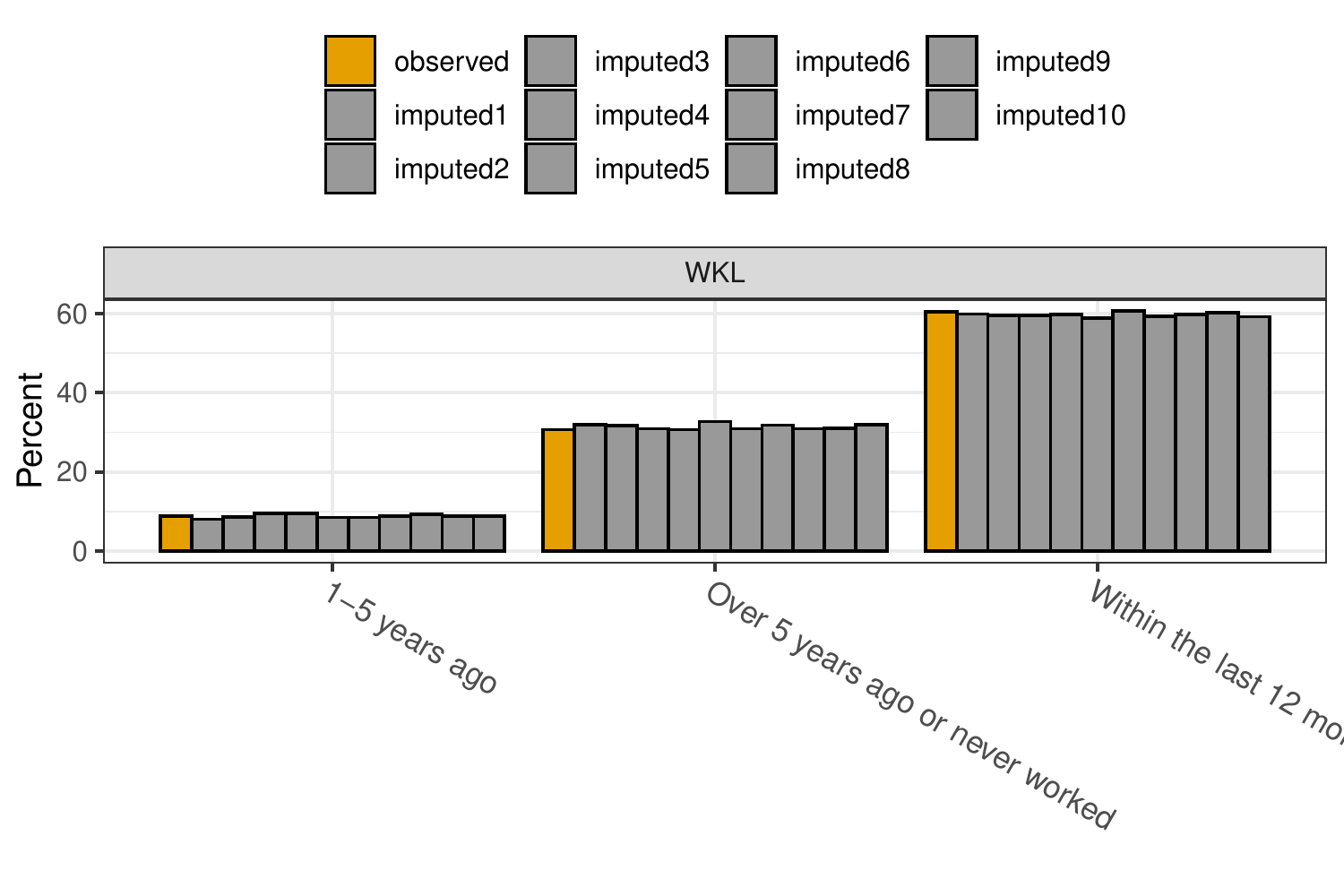}
	\caption{\label{MICEresults1}Marginal distribution of WKL from the observed data and each imputed dataset, using MICE. Barplots of the observed (yellow) and the 10 imputed (grey) are shown for the three levels of WKL. There is some variability across the imputed datasets. Overall they resemble the observed well.}
\end{figure}

A very common way to assess the quality of the imputations is to compare the estimated distributions in the observed and imputed datasets. We can compare the marginal distributions of any of the variables in the observed and imputed datasets, using the \code{marginal\_compare\_all\_imp} function. The function takes 3 arguments as input:
\begin{enumerate}
	\item \code{obsdata}, the observed data
	\item \code{impdata}, the list of m imputed datasets
	\item \code{vars}, the variable of interest.
\end{enumerate}
The output is a list containing:
\begin{enumerate}
	\item \code{Plot}, a barplot showing the marginal probability (as a percentage) of each level of the variable in the observed and imputed datasets
	\item \code{Comparison}, the table of the marginal probabilities (as a percentage) used to make the barplot.
\end{enumerate}
As an example, we can compare the marginal probability of each level of the variable WKL, in the observed and imputed datasets, for both MICE and the DPMPM engine, by using the following code. We load the \CRANpkg{tidyverse} library for making these plots.
\begin{verbatim}
library(tidyverse)
marginal_compare_all_imp(obsdata = X,
                         impdata = Imp_DPMPM$impdata,
                         vars = "WKL") 
marginal_compare_all_imp(obsdata = X,
                         impdata = Imp_MICE_reshape$impdata,
                         vars = "WKL") 
\end{verbatim}
The code creates the plots in Figures \ref{DPMPMresults1} and \ref{MICEresults1}.
For the most part in Figures \ref{DPMPMresults1} and \ref{MICEresults1}, both DPMPM and MICE result in point estimates from the imputed datasets that are very close to the observed data, which are to be expected under MCAR. There are no major noticeable differences between the two methods. The code can be applied in a similar manner to other variables, MAR and SEX, shown in the accompanying R file.
%

\subsection*{{\underline{Using the multiple imputation combining rules}}}
We now demonstrate how to use the combining rules to obtain single point estimates and corresponding 95\% confidence intervals for estimands of interest from all the imputed datasets. First, we compute the point estimates and corresponding standard errors for marginal and joint probabilities, from each imputed dataset, using the \code{compute\_probs} function. The function takes 2 arguments as input:
\begin{enumerate}
	\item \code{InputData}, the list of m imputed datasets
	\item \code{varlist}, a list of variable names (or combination of names) of interest.
\end{enumerate}
The output is a list of the marginal and/or joint probabilities in each imputed dataset. Next, we use the \code{pool\_estimated\_probs} function to pool the estimates from all the imputed datasets using the combining rules. The function takes 2 arguments as input:
\begin{enumerate}
	\item \code{ComputeProbsResults}, the output from the \code{compute\_probs} function
	\item \code{method}, the combining rules to use, where the options are "\code{imputation}", "\code{synthesis\_full}", "\code{synthesis\_partial}".
\end{enumerate}
The output is a list of tables containing the results after applying the combining rules. For example, suppose we are interested in estimating probabilities corresponding to (i) the marginal distribution of MAR, (ii) the marginal distribution of SEX, and (iii) the joint distribution of MAR and WKL, we can use the following code.
\begin{verbatim}
	varlist <- list(c("MAR"),c("SEX"),c("MAR","WKL")) #probabilities to evaluate
	prob_ex1_DPMPM <- compute_probs(InputData = Imp_DPMPM$impdata,
                                varlist = varlist)
pooledprob_ex1_DPMPM <- pool_estimated_probs(ComputeProbsResults = prob_ex1_DPMPM,
                                             method = "imputation")
                                             
	prob_ex1_MICE <- compute_probs(InputData = Imp_MICE_reshape$impdata,
                               varlist = varlist)
pooledprob_ex1_MICE <- pool_estimated_probs(ComputeProbsResults = prob_ex1_MICE,
                                            method = "imputation")
\end{verbatim}
When dealing with missing data imputation, the \code{method} must be set to "\code{imputation}". The first element of \code{pooledprob\_ex1\_DPMPM} for MAR is shown below, whereas the remaining output is omitted for brevity.
\begin{verbatim}
                      MAR Estimate   Std.Error       Df Statistic   CI_Lower   CI_Upper
1                Divorced   0.1083 0.011865562 9.000000  9.127254 0.08145823 0.13514177
2                 Married   0.5125 0.020191254 9.000001 25.382277 0.46682421 0.55817579
3 Never married or age<15   0.2953 0.018326210 9.000000 16.113534 0.25384323 0.33675677
4               Separated   0.0204 0.005678460 9.000000  3.592523 0.00755443 0.03324557
5                 Widowed   0.0635 0.008683588 9.000000  7.312645 0.04385636 0.08314364
\end{verbatim}


Each row represents the different levels of the corresponding variable(s). From left to right, the columns give the variable names and levels, the overall point estimates averaged across all imputed datasets, and the corresponding standard errors, degrees of freedom, test statistics, and confidence intervals.
Similarly, for \code{pooledprob\_ex1\_MICE}, we have
\begin{verbatim}
                      MAR Estimate   Std.Error Df Statistic    CI_Lower   CI_Upper
1                Divorced   0.1055 0.011501682  9  9.172571 0.079481387 0.13151861
2                 Married   0.5204 0.017156032  9 30.333355 0.481590360 0.55920964
3 Never married or age<15   0.2912 0.015438579  9 18.861839 0.256275507 0.32612449
4               Separated   0.0175 0.005031389  9  3.478165 0.006118207 0.02888179
5                 Widowed   0.0654 0.009065137  9  7.214452 0.044893235 0.08590676
\end{verbatim}
As the output shows, the results from both MICE and DPMPM are once again similar when looking at marginal probabilities of MAR, and both are indeed close to the results from the original sample without any missing data (which are excluded for brevity).

The \CRANpkg{NPBayesImputeCat} package also includes similar functions, \code{fit\_GLMs} and \code{pool\_fitted\_GLMs}, for fitting generalized linear models (GLMs) to each imputed datasets and pooling the results across all the datasets. The \code{fit\_GLMs}  function takes 2 arguments as input:
\begin{enumerate}
	\item \code{InputData}, the list of \code{m} imputed datasets
	\item \code{exp}, the GLM expression for the model of interest (for \code{nnet} which must be loaded first).
\end{enumerate}
The output is a list containing the estimated parameters from the GLM model fitted to each imputed dataset. The \code{pool\_fitted\_GLMs} pools the GLM estimates from all the imputed datasets using the combining rules. The function takes 2 arguments as input:
\begin{enumerate}
	\item \code{GLMResults}, the output from the \code{fit\_GLMs} function
	\item \code{method}, the combining rules to use, where the options are "\code{imputation}", "\code{synthesis\_full}", "\code{synthesis\_partial}".
\end{enumerate}
For example, to fit a multinomial logistic model of MAR on SEX, we can use the following code.
\begin{verbatim}
library(nnet)
model_ex1_DPMPM <- fit_GLMs(InputData = Imp_DPMPM$impdata,
                            exp = multinom(formula = MAR~SEX))
pool_fitted_GLMs(GLMResults = model_ex1_DPMPM,
                 method = "imputation")
\end{verbatim}
The second line yields the following output, with the numbers rounded up to 4 decimal places.
\begin{verbatim}
	                   Levels   Parameter Estimate Std.Error      Df Statistic CI_Lower CI_Upper
1                 Married (Intercept)   1.5431    0.1783  9.0033    8.6536   1.1398   1.9465
2                 Married     SEXMale   0.0282    0.2597  9.0150    0.1085  -0.5591   0.6155
3 Never married or age<15 (Intercept)  0.8719   0.1849  9.0034    4.7145   0.4536   1.2902
4 Never married or age<15     SEXMale  0.2626   0.2632  9.0139    0.9976  -0.3327   0.8578
5               Separated (Intercept)  -1.6225    0.4380  9.1346   -3.7044  -2.6111  -0.6339
6               Separated     SEXMale  -0.2124    0.7787 10.3909   -0.2728  -1.9386   1.5138
7                 Widowed (Intercept)  -0.0666    0.2174  9.0059   -0.3062  -0.5584   0.4252
8                 Widowed     SEXMale  -1.5906    0.4781  9.1614   -3.3265  -2.6693  -0.5118
\end{verbatim}

The \code{fit\_GLMs} and \code{pool\_fitted\_GLMs} functions perform a similar role to the \code{with} and \code{pool} functions in the \code{MICE} package. To fit the same model under MICE, we use the following code.
\begin{verbatim}
model_ex1_MICE <- with(data = Imp_MICE,
                       exp = multinom(formula = MAR~SEX))
summary(pool(model_ex1_MICE))
\end{verbatim}
The second line yields the following output, with the numbers rounded up to 4 decimal places.
\begin{verbatim}
                  y.level    term estimate std.error  statistic   df  p.value
1                 Married (Intercept)  1.6828 0.1835  9.1720  78.1913  0.0000
2                 Married     SEXMale -0.1735 0.2745 -0.6321  53.7652  0.5300
3 Never married or age<15 (Intercept)  0.8643 0.1962  4.4051  95.0509  0.0000
4 Never married or age<15     SEXMale  0.2828 0.2980  0.9491  48.8560  0.3473
5               Separated (Intercept) -1.5754 0.6032 -2.6117  19.3114  0.0170
6               Separated     SEXMale -0.7507 1.0821 -0.6937  17.6531  0.4969
7                 Widowed (Intercept)  0.1187 0.2481  0.4783  49.8076  0.6345
8                 Widowed     SEXMale -2.2404 0.6741 -3.3235  32.0590  0.0022
\end{verbatim}
The results are mostly similar, although we note that for most of the estimands, the standard errors are larger for MICE than the DPMPM engine. However, we also note that this illustration is based on data containing only $n=1000$ observations and $30\%$ missing data, so that differences in point estimates are not particularly surprising. Additional examples of fitting GLM models to the imputed datasets are shown in the accompanying R file.
%
%
%
%

\subsection{Multiple imputation for data with structural zeros}
\label{sec:illustrationsMI:zeros}

We now illustrate how to impute missing values for ACS sample 1 with 30\% missingness, using \CRANpkg{NPBayesImputeCat}, where there are structural zeros are present. Recall that the data is stored in the file, \file{ss16pusa\_sample\_zeros\_miss}. The general procedure is very similar to the one in the previous section where structural zeros are not present. However, we need to specify additional inputs to account for the structural zeros, when generating the imputed datasets. Once the imputed datasets have been created, the utility functions used to computed sample estimates and pool them using the combining rules, are exactly the same as before. First, we begin by creating \code{MCZ}, the data frame containing the structural zeros definition.

\subsection*{{\underline{Create a file to store structural zeros cases}}}
Previously when there are no structural zeros, \code{MCZ} is set to \code{NULL} . Here when there are structural zeros cases in the application, one should write the \code{MCZ} data frame, following two general rules:

\begin{enumerate}
	\item Variables in \code{MCZ} must be factors with the same levels as the original data.
	\item Placeholder components are represented with NAs.
\end{enumerate}
The script below is a sample script to store the structural zeros definition shown in Table \ref{tab:ACSzeros}.
\begin{verbatim}
AGEP <- c(16, 16, 16, 16, 17, 17, 17, 17)
SCHL <- c("Bachelor's degree", "Doctorate degree", "Master's degree", "Professional degree", 
          "Bachelor's degree", "Doctorate degree", "Master's degree", "Professional degree")
MAR <- rep(NA, 8)
SEX <- rep(NA, 8)
WKL <- rep(NA, 8)
MCZ <- as.data.frame(cbind(AGEP, MAR, SCHL, SEX, WKL))
\end{verbatim}

First, we create a vector of AGEP consisting of 4 replicates of value 16 and 4 replicates of value 17, and a vector of SCHL consisting the degree types which induce structural zeros cases with AGEP. Second, we create vectors of MAR, SEX, and WKL, each is a vector of length 8, with each element being NA. These are placeholder components, and since the structural zeros cases do not involve these three variables, all elements are NAs. Third, we need to create a data frame using \code{as.data.frame} and \code{cbind}. It is necessary to input the variables in the same order as in the original data (the order of variables in Table \ref{tab:ACSvars1}). We save the data frame \code{MCZ} for later use.

\subsection*{{\underline{Load the sample}}}
Now, we load the sample data and make sure that all variables are unordered factors.
\begin{verbatim}
data("ss16pusa_sample_zeros_miss")
X <- ss16pusa_sample_zeros_miss
p <- ncol(X)
for (j in 1:p){
  X[,j] <- as.factor(X[,j])
  MCZ[,j] <- factor(MCZ[,j], levels = levels(X[,j]))
}
\end{verbatim}

\subsection*{{\underline{Generate the imputed datasets}}}
Initializing the DPMPM engine follows the exact same approach as before. The only difference is that we can now supply the two arguments specific to structural zeros, that is, \code{MCZ} and \code{Nmax}. That is, to initialize, we can run the following code.
\begin{verbatim}
model <- CreateModel(X = X,
                     MCZ = MCZ,
                     K = 30,
                     Nmax = 20000,
                     aalpha = 0.25,
                     balpha = 0.25,
                     seed = 521)
\end{verbatim}
As before, we select \code{K} based on initial runs. We now also do the same for \code{Nmax}. If the value of either always hits the set values, we should re-run the model by specifying larger values, to allow for a large enough number of mixture components and augmented observations, to cluster the observations appropriately. As before, we can also save and track posterior samples of the parameters in the sampler using the \code{EnableTracer} option. Sample scripts are included in the accompanying R file. 

After setting \code{K} and \code{Nmax}, we can now run the DPMPM imputation engine with structural zeros to create $m$ imputed datasets. The function \code{DPMPM\_zeros\_imp} takes 12 arguments as input: 
\begin{enumerate}
	\item \code{X}, the original data with missing values
	\item \code{MCZ}, data frame containing the structural zeros definition
	\item \code{Nmax}, an upper truncation limit for the augmented sample size
	\item \code{nrun}, the number of MCMC iterations
	\item \code{burn}, the number of burn-in
	\item \code{thin}, the number of thinning
	\item \code{K}, the maximum number of mixture components (i.e. the maximum number of latent classes in the DPMPM imputation engine) \item \code{aalpha}, the hyper parameter $a_{\alpha}$ in stick-breaking prior distribution in Equation (\ref{prior-alpha})
	\item \code{balpha}, the hyper parameter $b_{\alpha}$ in stick-breaking prior distribution in Equation (\ref{prior-alpha})
	\item \code{m}, the number of imputations
	\item \code{seed}, the seed value
	\item \code{silent}, default to TRUE. Set this parameter to FALSE if more iteration info are to be printed.
\end{enumerate}
The output of \code{DPMPM\_zeros\_imp} is similar to the output of \code{DPMPM\_nozeros\_imp}, 
except that now it also includes \code{Nmis}, the saved posterior draws of the augmented sample size, which can be used to check MCMC convergence.

To run the \code{DPMPM\_zeros\_imp} function to impute missing data for ACS sample 1 with 30\% missingness, we run the code below. For this demonstration, we set \code{Nmax} to 200000, \code{nrun} to 10000, \code{burn} to 5000, \code{thin} to 50, \code{K} to 80,  both \code{aalpha} and \code{balpha} 0.25, and \code{m} to 10. Finally, we set the \code{seed} to 653.
\begin{verbatim}
m <- 10
Imp_DPMPM <- DPMPM_zeros_imp(X = X,
                             MCZ = MCZ,
                             Nmax = 200000,
                             nrun = 10000,
                             burn = 5000,
                             thin = 50,
                             K = 80,
                             aalpha = 0.25,
                             balpha = 0.25,
                             m = m,
                             seed = 653,
                             silent = TRUE)
\end{verbatim}
As before, it is straightforward to run MCMC diagnostics on the tracked elements of \code{Imp\_DPMPM}. Also, \code{Imp\_DPMPM} contains the list of the imputed datasets. 
Analysts then can compute sample estimates for estimands of interest in each imputed dataset, and assess their quality or also combine them using the combining rules, by using all the same functions as before. That is, using \code{marginal\_compare\_all\_imp}, \code{compute\_probs}, \code{pool\_estimated\_probs}, \code{fit\_GLMs} and \code{pool\_fitted\_GLMs}. Examples are included in the accompanying R file. 

Currently, there are no direct options in the \CRANpkg{mice} package to incorporate structural zeros. Therefore, we do not explore comparisons with the MICE engine for data containing structural zeros.

\section{Synthetic data applications}\label{sec:illustrationsSD}

Without loss of generality, suppose we want to generate partially synthetic datasets for the ACS sample 2 (\file{ss16pusa\_sample\_nozeros}), where no structural zeros are present. \CRANpkg{NPBayesImputeCat} includes functionality to generate fully synthetic data as well, but we exclude its illustration for brevity.
The \CRANpkg{NPBayesImputeCat} package currently does not accommodate data synthesis with structural zeros.

We also implement a popular synthetic data generation method, CART (using the \CRANpkg{synthpop} package in R), to ACS sample 2 and compare the results \citep{Reiter2005CART} to DPMPM. 

\subsection*{{\underline{Load the sample data}}}

First, we load the sample data, the ACS sample 2, and make sure that all variables are set as factors.

\begin{verbatim}
data(ss16pusa_sample_nozeros)
X <- ss16pusa_sample_nozeros
p <- ncol(X)
for (j in 1:p){
  X[,j] <- as.factor(X[,j])
}
\end{verbatim}

\subsection*{{\underline{Generate the synthetic datasets}}}

Initializing the DPMPM synthesizer follows the exact same approach as before for the missing data imputation applications. We run the following code.

\begin{verbatim}
model <- CreateModel(X = X,
                     MCZ = NULL,
                     K = 80,
                     Nmax = 0,
                     aalpha = 0.25,
                     balpha = 0.25,
                     seed = 973)
\end{verbatim}

After setting \code{K} based on the initial runs, we now run the DPMPM synthesizer without structural zeros to create \code{m} synthetic datasets. The function \code{DPMPM\_nozeros\_syn} takes 12 arguments as input:

\begin{enumerate}
	\item \code{X}, the original data with missing values
	\item \code{dj}, the vector recording the number of categories of the variables
	\item \code{nrun}, the total number of MCMC iterations
	\item \code{burn}, the number of burn-ins
	\item \code{thin}, the number of thinnings
	\item \code{K}, the maximum number of mixture components
	\item \code{aalpha}, the hyper parameter $a_{\alpha}$ in stick-breaking prior distribution in Equation (\ref{prior-alpha})
	\item \code{balpha}, the hyper parameter $b_{\alpha}$ in stick-breaking prior distribution in Equation (\ref{prior-alpha})
	\item \code{m}, the number of synthetic datasets
	\item \code{vars}, the names of the variables to be synthesized
	\item \code{seed}, the seed value
	\item \code{silent}, default to TRUE. Set this parameter to FALSE if more iteration info are to be printed.
\end{enumerate}

The output of \code{DPMPM\_nozeros\_syn} is a list containing: 

\begin{enumerate}
	\item \code{syndata}, the list of $m$ synthetic datasets
	\item \code{origdata}, the original data \code{X}
	\item \code{alpha}, the saved draws of $\alpha$, which can be used to check MCMC convergence
	\item \code{kstar}, the saved numbers of occupied mixture components, which can be used to check MCMC convergence and track whether the upper bound \code{K} is set large enough.
\end{enumerate}

To run the \code{DPMPM\_nozeros\_syn} function to generate synthetic data for ACS sample 2, we run the code below. For this demonstration, we create partially synthetic data where marital status (MAR) and when last worked (WKL) are synthesized, and we set \code{nrun} to 10000, \code{burn} to 5000, \code{thin} to 50, \code{K} to 80, both \code{aalpha} and \code{balpha} to 0.25, and \code{seed} to 837. Recall that \code{dj} stores the vector of levels of the variables, which are 5 for MAR, 2 for SEX, and 3 for WKL.
\begin{verbatim}
dj <- c(5, 2, 3)
m <- 5
Syn_DPMPM <- DPMPM_nozeros_syn(X = X,
                               dj = dj,
                               nrun = 10000,
                               burn = 5000,
                               thin = 50,
                               K = 80,
                               aalpha = 0.25,
                               balpha = 0.25,
                               m = 5,
                               vars = c("MAR", "WKL"),
                               seed = 837,
                               silent = TRUE)
\end{verbatim}

MCMC diagnostics can be run based on the tracked elements of \code{Syn\_DPMPM}. To access the synthetic datasets one at a time, we do the following.

\begin{verbatim}
syndata3 <- Syn_DPMPM$syndata[[3]] #for the third synthetic dataset
\end{verbatim}

Analysts then can compute sample estimates for estimands of interest in each synthetic dataset, and combine them using the combining rules. For comparison, we use the \CRANpkg{synthpop} package to generate synthetic data for the same dataset with CART synthesizer. The default synthesizer for \CRANpkg{synthpop} is CART, and the \code{visit.sequence} input argument allows the users to indicate which variables to be synthesized. To match with what we have done with the \CRANpkg{NPBayesImputeCat}, we set MAR and WKL for \code{visit.sequence}  to generate partially synthetic data.

\begin{verbatim}
library(synthpop)
m <- 5 
Syn_CART <- syn(data = X,
                m = 5,
                seed = 123,
                visit.sequence = c("MAR", "WKL")) 
\end{verbatim}

Next, we demonstrate how to access the utility of the synthetic datasets from the two methods, and also use the combining rules.

\subsection*{{\underline{Assess utility of the synthetic datasets}}}

Similar to what we have introduced for assessing the quality of the imputations, for synthesis we compare the marginal distributions of any of the variables in the observed and imputed datasets, using the \code{marginal\_compare\_all\_syn} function. The function takes 3 arguments as input:

\begin{enumerate}
	\item \code{obsdata}, the observed data
	\item \code{syndata}, the list of m synthetic datasets
	\item \code{vars}, the variable of interest.
\end{enumerate}
The output is a list containing:
\begin{enumerate}
	\item \code{Plot}, a barplot showing the marginal probability (as a percentage) of each level of the variable in the observed and synthetic datasets
	\item \code{Comparison}, the table of the marginal probabilities (as a percentage) used to make the barplot
\end{enumerate}

\begin{figure}[h!]
	\centering
	\includegraphics[width=0.7\linewidth]{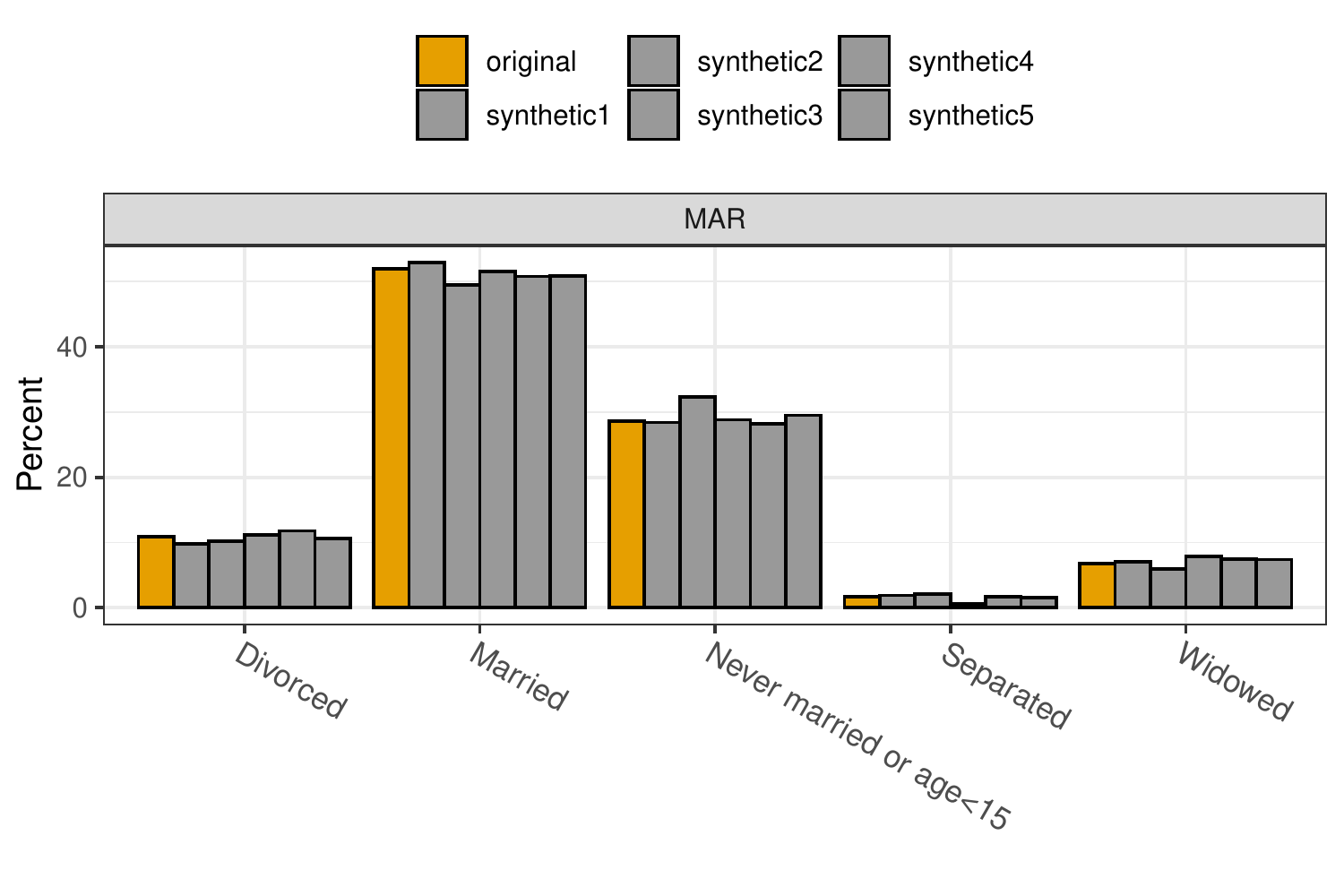}
	\caption{\label{DPMPMresults3}Marginal distribution of MAR from the observed data and each synthetic dataset, using DPMPM. Barplots of the original (yellow) and the 5 synthetic (grey) are shown for the five levels of MAR. There is some variability across the synthetic datasets. Overall they resemble the original well.}
\end{figure}

\begin{figure}[h!]
	\centering
	\includegraphics[width=0.7\linewidth]{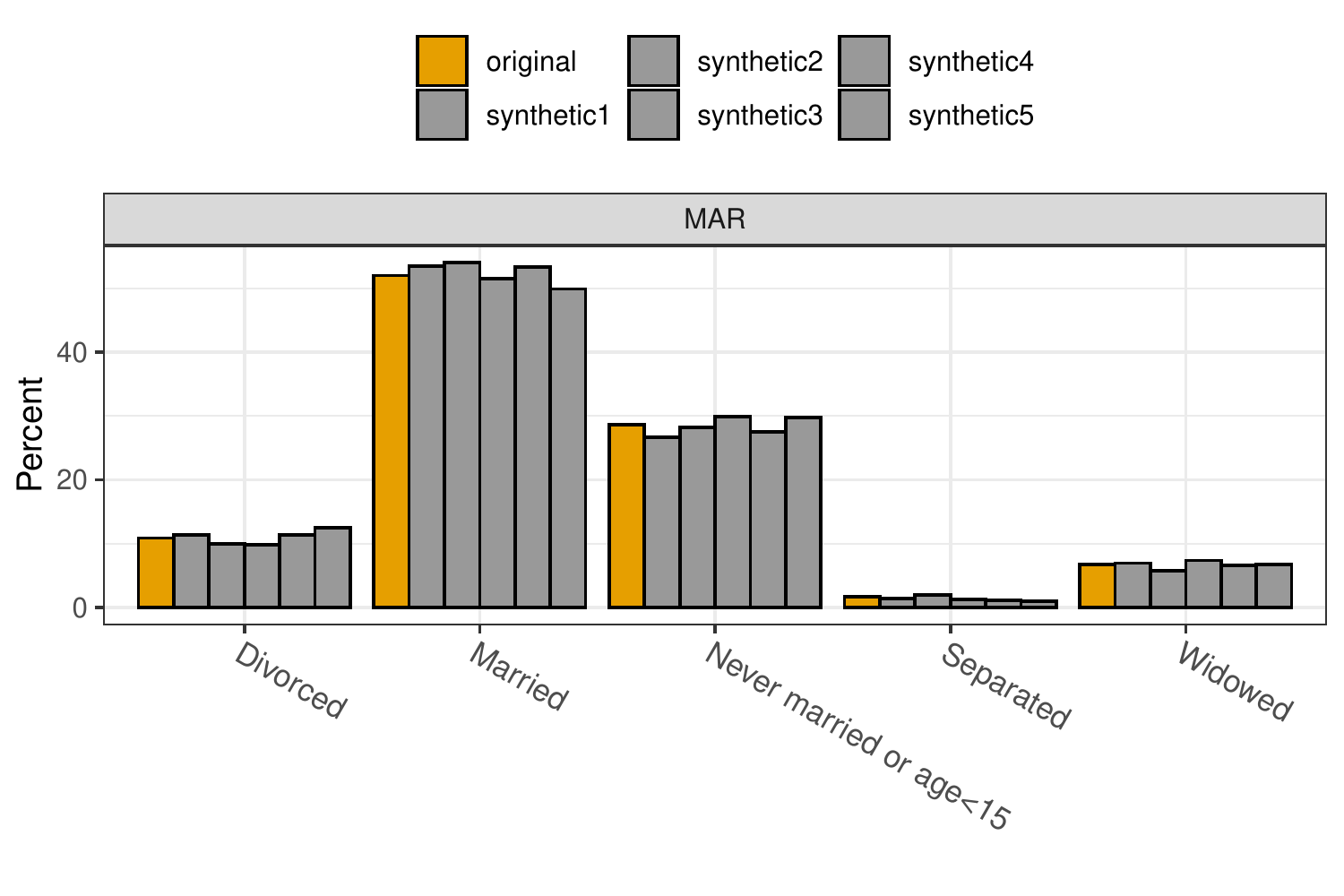}
	\caption{\label{CARTresults}Marginal distribution of MAR from the observed data and each synthetic dataset, using CART. Barplots of the original (yellow) and the 5 synthetic (grey) are shown for the five levels of MAR. There is some variability across the synthetic datasets. Overall they resemble the original well.}
\end{figure}

The following code compares the marginal probability of each level of MAR.
\begin{verbatim}
marginal_compare_all_syn(obsdata = X,
                         syndata = Syn_DPMPM$syndata,
                         vars = "MAR")
marginal_compare_all_syn(obsdata = X,
                         syndata = Syn_CART$syn,
                         vars = "MAR")

\end{verbatim}
The code creates the plots in Figures \ref{DPMPMresults3} and \ref{CARTresults}.


For the most part in Figures \ref{DPMPMresults3} and \ref{CARTresults}, both DPMPM and CART result in point estimates from the synthetic datasets that are very close to the observed data. There are no major noticeable differences between the two methods. The code can be applied in a similar manner to other variables and examples are included in the accompanying R file. 

%

\subsection*{{\underline{Using the synthetic data combining rules}}}

We now demonstrate how to use the combining rules to obtain single point estimates and corresponding 95\% confidence intervals for estimands of interest from all the synthetic datasets. Similar to what we have done with the multiple imputation combining rules previously, we use the same set of functions: \code{compute\_probs} and \code{pool\_estimated\_prob}.

For example, suppose we are interested in estimating probabilities corresponding to (i) the marginal distribution of MAR, (ii) the marginal distribution of SEX, (iii) the marginal distribution of WKL, and (iv) the joint distribution of MAR and WKL, we can use the following code:
\begin{verbatim}
varlist <- list(c("MAR"), c("SEX"), c("WKL"), c("MAR","WKL")) #probabilities to evaluate
prob_ex1_DPMPM <- compute_probs(InputData = Syn_DPMPM$syndata,
                                varlist = varlist)
pooledprob_ex1_DPMPM <- pool_estimated_probs(ComputeProbsResults = prob_ex1_DPMPM,
                                             method = "synthesis_partial")
                                             
prob_ex1_CART <- compute_probs(InputData = Syn_CART$syn,
                               varlist = varlist)
pooledprob_ex1_CART <- pool_estimated_probs(ComputeProbsResults = prob_ex1_CART,
                                            method = "synthesis_partial")
\end{verbatim}
As noted before, when dealing with partially synthetic data, the \code{method} must be set to ``synthesis\_partial''. We only include the first element of \code{pooledprob\_ex1\_DPMPM} for MAR for brevity. 
\begin{verbatim}
                      MAR Estimate   Std.Error        Df Statistic    CI_Lower   CI_Upper
1                Divorced   0.1072 0.010406613 293.63208 10.301142 0.086718995 0.12768100
2                 Married   0.5112 0.016738961 338.91020 30.539531 0.478274660 0.54412534
3 Never married or age<15   0.2944 0.016234667  88.41528 18.134034 0.262139123 0.32666088
4               Separated   0.0158 0.004718411  43.64386  3.348585 0.006288476 0.02531152
5                 Widowed   0.0714 0.008826970 178.61153  8.088846 0.053981434 0.08881857
\end{verbatim}


Each row represents the different levels of the corresponding variable(s). From left to right, the columns give the variable names and levels, the overall point estimates averaged across all imputed datasets, and the corresponding standard errors, degrees of freedom, test statistics, and confidence intervals.
Similarly, for \code{pooledprob\_ex1\_CART}, we have
\begin{verbatim}
                      MAR Estimate   Std.Error       Df Statistic    CI_Lower   CI_Upper
1                Divorced   0.1104 0.011044293 104.5372  9.996113 0.088500078 0.13229992
2                 Married   0.5244 0.017530887 111.6932 29.912919 0.489663753 0.55913625
3 Never married or age<15   0.2842 0.015588823 149.5745 18.231011 0.253397249 0.31500275
4               Separated   0.0138 0.004054405 134.0083  3.403705 0.005781098 0.02181890
5                 Widowed   0.0672 0.008348413 392.0400  8.049434 0.050786740 0.08361326
\end{verbatim}


As the output shows, the results from both CART and DPMPM are once again similar when looking at marginal probabilities of MAR, and both are indeed close to the results from the original sample (excluded for brevity).

Lastly, we demonstrate the use of \code{fit\_GLMs} and \code{pool\_fitted\_GLMs}, for fitting generalized linear models (GLMs) to each synthetic datasets and pooling the results across all the datasets. For example, to fit a logistic model of SEX given MAR and WKL, we can use the following code.
\begin{verbatim}
model_ex1_DPMPM <- fit_GLMs(InputData = Syn_DPMPM$syndata,
                            exp = glm(formula = SEX~WKL+MAR,
                                      family = binomial))
pool_fitted_GLMs(GLMResults = model_ex1_DPMPM,
                 method = "synthesis_partial")
\end{verbatim}
The second line yields the following output, with the numbers rounded up to 4 decimal places.
\begin{verbatim}
                                       Estimate Std.Error        Df  Statistic    CI_Lower   CI_Upper
(Intercept)                         -0.5954 0.3548  44.1182 -1.6781 -1.3105  0.1196
WKLOver 5 years ago or never worked  0.0946 0.2823 179.2380  0.3351 -0.4625  0.6517
WKLWithin the last 12 months         0.3629 0.3042  33.3456  1.1929 -0.2558  0.9816
MARMarried                           0.3517 0.2581  48.6246  1.3625 -0.1671  0.8706
MARNever married or age<15           0.4293 0.2560 119.8199  1.6771 -0.0775  0.9360
MARSeparated                         0.0710 0.6589 545.7313  0.1078 -1.2232  1.3652
MARWidowed                          -1.0156 0.3864 543.1589 -2.6285 -1.7746 -0.2566
\end{verbatim}

We use the following code to fit the same model under CART.
\begin{verbatim}
model_ex1_CART <- fit_GLMs(InputData = Syn_CART$syn,
                           exp = glm(formula = as.factor(SEX)~WKL+MAR,
                                     family = binomial))
pool_fitted_GLMs(GLMResults = model_ex1_CART,
                 method = "synthesis_partial")
\end{verbatim}
The second line yields the following output, with the numbers rounded up to 4 decimal places.
\begin{verbatim}
                                               Estimate  Std.Error         Df   Statistic   CI_Lower  CI_Upper
(Intercept)                         -0.0292 0.3155  103.8094 -0.0927 -0.6549 0.5964
WKLOver 5 years ago or never worked -0.0610 0.2941   56.6928 -0.2074 -0.6499 0.5279
WKLWithin the last 12 months        -0.1309 0.2602  125.4832 -0.5032 -0.6460 0.3841
MARMarried                           0.0316 0.2151 3096.9978  0.1471 -0.3902 0.4534
MARNever married or age<15           0.0230 0.2401  319.0640  0.0956 -0.4495 0.4954
MARSeparated                         0.3378 0.6530  187.2752  0.5172 -0.9505 1.6260
MARWidowed                           0.1148 0.3365  477.2065  0.3412 -0.5464 0.7760
\end{verbatim}

The output of fitting this GLM to the original sample is used as the benchmark for our utility evaluation.
\begin{verbatim}
                                    Estimate Std. Error z value Pr(>|z|)    
(Intercept)                          -0.5937     0.2897  -2.050 0.040395 *  
WKLOver 5 years ago or never worked   0.2859     0.2539   1.126 0.260163    
WKLWithin the last 12 months          0.5054     0.2358   2.144 0.032065 *  
MARMarried                            0.1380     0.2129   0.649 0.516647    
MARNever married or age<15            0.3994     0.2273   1.757 0.078965 .  
MARSeparated                         -0.3538     0.5463  -0.648 0.517183    
MARWidowed                           -1.3673     0.3867  -3.536 0.000406 ***
\end{verbatim}

The utility results are different between DPMPM and CART, and we note that for most of the estimands, the DPMPM produces more accurate estimation than the CART, when compared to the results from the original sample. These indicate that for this particular sample, a joint model fitted by the DPMPM does a better job preserving relationships among variables, compared to a series of conditional model fitted by the CART. Additional examples of fitting GLM models to the synthetic datasets are shown in the accompanying R file.
%

\section{Concluding remarks} \label{sec:summary}

In this paper, we have presented the DPMPM models for multivariate categorical data, and illustrations of using the \CRANpkg{NPBayesImputeCat} package for multiple imputation and synthetic data applications. Users can take the output and extract imputed and synthetic datasets, then conduct statistical analyses of their choice and use the appropriate combining rules to obtain valid estimates. Interested readers can refer to the package documentation for additional features.

While the \CRANpkg{NPBayesImputeCat} package has been developed primarily for multiple imputation and synthetic data purposes, users can also use it for DPMPM model estimation. For example, following the illustrations for synthetic data, a data analyst is able to obtain parameter draws of several key parameters from the MCMC chain: i) the DP concentration parameter $\alpha$, ii) the mixture probability vectors $\{\pi_k\}$, and iii) the Multinomial probability vectors $\{\theta_k^{(j)}\}$. The analyst can then further conduct analyses of the clustering of the observations in the MCMC chain, and other questions of interest.

Users can report bugs at our GitHub repo: \url{https://github.com/monika76five/NPBayesImputeCat}.

\bibliography{hu-akande-wang}

\address{Jingchen Hu\\
	Mathematics and Statistics Department\\
	Vassar College\\
	Box 27, 124 Raymond Ave \\
	Poughkeepsie, NY 12604, USA\\
	E-mail: \email{jihu@vassar.edu}}

\address{Olanrewaju Akande\\
	Social Science Research Institute, and\\
	The Department of Statistical Science\\
	Box 90989, Duke University\\
	Durham, NC 27708, USA\\
	E-mail: \email{olanrewaju.akande@duke.edu}}

\address{Quanli Wang\\
	Department of Statistical Science\\
	Box 90251, Duke University\\
	Durham, NC 27708, USA\\
	E-mail: \email{quanliwang20@gmail.com}}